# The Schrodinger-Chetaev Equation in Bohmian Quantum Mechanics and Diffusion Mechanism for Alpha Decay, Cluster Radioactivity and Spontaneous Fission

V.D. Rusov[1], S. Cht. Mavrodiev[2], M.A. Deliyergiyev[1]

[1]*Department of Theoretical and Experimental Nuclear Physics*

*Odessa National Polytechnic University, Ukraine*

[2] *The Institute for Nuclear Research and Nuclear Energy, BAS, Sofia, Bulgaria*

**Abstract**

In the framework of Bohmian quantum mechanics supplemented with the Chetaev theorem on stable trajectories in dynamics in the presence of dissipative forces we have shown the possibility of the classical (without tunneling) universal description of radioactive decay of heavy nuclei, in which under certain conditions so called noise-induced transition is generated or, in other words, the stochastic channel of alpha decay, cluster radioactivity and spontaneous fission conditioned by the Kramers diffusion mechanism.

Based on the ENSDF database we have found the parametrized solutions of the Kramers equation of Langevin type by Alexandrov dynamic auto-regularization method (FORTRAN program REGN-Dubna). These solutions describe with high-accuracy the dependence of the half-life (decay probability) of heavy radioactive nuclei on total kinetic energy of daughter decay products.

The verification of inverse problem solution in the framework of the universal Kramers description of the alpha decay, cluster radioactivity and spontaneous fission, which was based on the newest experimental data of alpha-decay of even-even super heavy nuclei (Z=114, 116, 118) have shown the good coincidence of the experimental and theoretical half-life depend upon of alpha-decay energy.

[1] Corresponding author: Prof. Rusov V.D., Head of Department of Theoretical and Experimental Nuclear Physics, Odessa National Polytechnic University, Shevchenko ave. 1, Odessa, 65044, Ukraine

Fax: + 350 482 641 672, E-mail: siiis@te.net.ua

# 1. Introduction

From the time of its discovery the atomic nucleus were used for testing the new physics ideas, for example, like tunneling [1, 2], superfluidity and superconductivity [3], Josephson nuclear effect [4], $\pi$ - condensate [5], dynamical supersymmetry [6] and nuclear quantum phase transition [7], quantum [8], dynamical and constructive [9] chaos, nuclear stochastic resonance [9]. In this sense the nucleus is undoubtedly the essentially nonlinear dynamical system, which can be effectively used for investigating different nonlinear effects on extremely small (nuclear) scales.

In this paper we propose an alternative model of heavy radioactive nuclear decay, which is based on the classical jump of daughter particle over the nuclear potential barrier as a result of diffusion induced by noise and not on the traditional quantum effect of "tunneling" or "percolation" through this barrier. The idea of such description can be naturally and clearly formalized in language of Bohmian quantum mechanics [10, 11] supplemented with the Chetaev generalized theorem on stable trajectories in classical dynamics in the presence of dissipative forces [12]. It became possible, when we obtained the generalized stability condition for Hamiltonian holonomic systems as Schrodinger equation in the framework of classical mechanics [12, 13]

$$i\hbar \frac{\partial \psi}{\partial t} = -\frac{\hbar^2}{2m}\Delta\psi + U\psi, \qquad (1)$$

using the Chetaev theorem [14]. We will call below this equation as Schrodinger-Chetaev equation to emphasize in that way a feature of its origin.

At the same time, it was shown that the wave function $\psi$, which describes the real electromagnetic waves of "de Broglie" type [12, 13]

$$\psi = A\exp\left(\frac{i}{\hbar}S\right), \qquad (2)$$

obeys the following condition

$$\frac{\partial A}{\partial t} = -\nabla A \frac{\nabla S}{m} \qquad (3)$$

and by virtue of Eq. (3) is the reason of dissipative forces origin [12-14], whose dissipation energy is equal to the Bohm's quantum potential [8, 9] and looks like

$$Q = -\frac{\hbar^2}{2m}\frac{\Delta A}{A}, \qquad (4)$$

where $S$ is action; $h = 2\pi\hbar$ is Plank constant; $A$ is amplitude, which in the general case is real function of the coordinates $q_i$ and time $t$.

Here the natural question arises: "What is the real reason of dissipative forces origin?" As is ascertained in Refs.[12, 13] the main reason is the translational precession of the particle spin. This notion was fist proposed by the Polish physicist Grysinsky, who has shown that, using real de Broglie waves in the form of oscillating electromagnetic field of photon or electron caused by translational precession of the spin, it is possible to explain the particle interference and diffraction phenomena in the framework of Newtonian mechanics and classical electrodynamics [15,16]. Using the concept of localized electron, i.e. in the framework of classical dynamics, he has also shown how electron spin axis precession hides behind the Sommerfeld quantization in-tegral, and how alternating electromagnetic field caused by the precession of its magnetic axis hides behind the wave field of electron [15,16].

Note, that the notion of translation precession of the particle spin is natural in the view of the Chetaev's theorem on stable trajectories in dynamics. In this sense, it is easy to see that at $\partial A/\partial t=0$ the translational precession of the spin around particle magnetic axis is just such gyroscopic force, which provide automatically the satisfaction of special requirements to the structure of items in Eq. (3) and, thereby, ensures the particle stable motion of zitterbewegung type [11, 17], which was, apparently, really observed in the experiments of Catillon et al [18]. In other words, the Gryzinski assumption of the translational precession of the spin generating de Broglie electromagnetic disturbing wave [15] follows naturally at $\partial A/\partial t=0$ from conditions (3) of the generalized Chetaev's theorem on the stable trajectories in dynamics [14].

On the other hand, it is well known that in general case the dissipation is defined by friction and random Langevin force with zero average value on the corresponding space-time sampling. Then force balance equation in one-dimensional case has the form:

$$\frac{\partial Q}{\partial x} = F_{frict}(x,\dot{x}) + F_L(x,t), \quad \text{where} \quad \langle F_L \rangle = 0. \qquad (5)$$

Note, that this equation is a consequence of the fact that the force of friction $F_{frict}$ as well as the Langevin force $F_L$ is produced by unified source, i.e. by the interaction of particle with environment, for example, with heat reservoir.

Now let us write down the Hamilton-Jacoby quantum equation, which is easily to obtain by the substitution of wave function (2) into the Schrodinger-Chetaev equation (1) and posterior isolation of imaginary part:

$$\frac{\partial S}{\partial t} = -\left[\frac{(\nabla S)^2}{2m} + U + Q\right]. \qquad (6)$$

Then, if Hamiltonian of the system is not explicitly time-dependent, we obtain the well known Langevin equation [19,20] for one-dimensional case by differentiation of Eq. (6) with respect to coordinate $x$ and allowance for Eq. (5):

$$m\ddot{x} = -dU_x - F_{frict}(x,\dot{x}) + F_L(x,t). \qquad (7)$$

It is important to note here, that the notion "dissipation" for any physical system is practically always connected with a separation of the degrees of freedom of the system on collective (slow) and intrinsic (fast, over which averaging is realized). At the same time the dynamics of the collective variables in this case is similar to the Brownian particle dynamics because the change of particle energy in one interaction with a thermostat is very small. Adequacy of such description is based on the assumption that time for the attainment of thermal equilibrium in the system with intrinsic degrees of freedom is considerably smaller than typical time scales of collective motion connected with Brownian particle dynamics. As basic equation of such diffusion models, which describe the system evolution as the wandering of virtual Brownian particle in the space of collective variables, it is convenient to use the Langevin stochastic equation (7).

It is possible to show [21-23] that the time evolution of the Brownian particle with mass $m$, which is in a potential well $U(r, t)$ and interacts with equilibrium thermal reservoir, can be described by the following concrete Langevin stochastic equation:

$$m\ddot{r} = -dU_r(r,t) - \gamma(r)\dot{r} + \sqrt{D(r)}\Gamma(t), \qquad (8)$$

where the intensity $D(r)$ of the Langevin force $\Gamma(t)$ is connected with friction coefficient $\gamma(r)$ and temperature $T$ of reservoir by well-known relation

$$D(r) = \gamma(r) k_B T, \qquad (9)$$

which characterizes a special case of fluctuation-dissipation theorem [21-23], and the statistic properties of random force $\Gamma(t)$ are determined in the following way:

$$\langle \Gamma(t) \rangle = 0; \quad \langle \Gamma(t) \Gamma(t') \rangle = \delta_\varepsilon (t - t'), \qquad (10)$$

where $\delta_\varepsilon(t-t')$ is a "smeared-out $\delta$- function" with a range $\varepsilon$, e.g. represented by expression

$$\delta_\varepsilon (t - t') = \begin{cases} 1/\varepsilon, & \text{for } -\varepsilon/2 \le t - t' \le \varepsilon/2, \\ 0 & \text{otherwise} \end{cases}. \qquad (11)$$

Assuming that friction coefficient $\gamma$ is coordinate-independent as before and using expression for the momentum of the collective motion of physical system $p = m\dot{r}$, we can represent the Langevin equation (8) for the canonically conjugate variables $\{p, r\}$ as:

$$\dot{r} = p/m, \qquad (12)$$

$$\dot{p} = -\frac{dU}{dr} - \frac{\gamma}{m} p + \sqrt{D}\Gamma(t). \qquad (13)$$

When $U = 0$, the solution of Eqs. (12)-(13) is well-known analytical results for the average values $<p^2>$ and $<r^2>$ for free Brownian particle. When $U \ne 0$, we return to the well-known Kramers diffusion problem of stochastic transitions over the potential barrier with velocity, which in the general case is determined (taking into account Eq. (9)) by the following relation [24,25]:

$$w_K = \frac{\omega_{\min}}{2\pi} \left\{ \left[ 1 + \left( \frac{\beta}{2\omega_{\max}} \right)^2 \right]^{1/2} - \frac{\beta}{2\omega_{\max}} \right\} \exp\left( -\frac{\Delta U}{\langle \varepsilon \rangle} \right), \quad \beta = \frac{\gamma}{m} \ge \frac{\omega_{\max}}{10}, \qquad (14)$$

where the average energy of the harmonic oscillator (oscillating charge) motion under the action of a thermal and zero-point (at $T=0$) radiation is equal [26]:

$$\langle \varepsilon \rangle = \frac{\hbar \omega_{\min}}{2} \coth\left( \frac{\hbar \omega_{\min}}{2 k_B T} \right) = \begin{cases} k_B T & \text{for high } T, \\ \hbar \omega_{\min}/2, & \text{when } T \to 0 \end{cases},$$

$\omega_{min}$ and $\omega_{max}$ are the angular frequencies of potential $U(x)$ in the potential minimum and in the vertex of barrier, respectively; $\Delta U$ is potential barrier height; $k_B$ is Boltzmann constant.

In case of induced decays the nuclear temperature ($T$) sets approximately (for the given density of simple-particle levels $g$) the number $n = gT$ of one-particle levels, which make contribution to the nuclear level density. At the same time the thermodynamic temperature by itself is linked through the parameter of level density $a = \pi^2 g/6$ with the internal excitation energy $E^*$ by well-known relation of Fermi-gas model [21-23]:

$$T = (E^*/a)^{1/2}, \quad MeV.$$

In general case, i.e. for the universal description of induced and spontaneous decays (at $T=0$), we use this relation in the following form:

$$\langle \varepsilon \rangle = (E^*/a)^{1/2}. \tag{15}$$

Note for further calculations that according to the experimental data of level density parameter $a$ the $A/a$ ratio for nuclei with mass number A is equal to $(8\pm1)$ MeV [22, 23].

Now we are ready for the description of our subject of interest, i.e. for investigation of the Kramers stochastic transitions over the potential barrier (14) in nonlinear nuclear dynamics for α- decay, cluster radioactivity and spontaneous fission.

## 2. The Kramers's channel of α-decay, cluster radioactivity and spontaneous fission

Let us consider the general case of a potential, in which some nuclear particle, for example, α-particle, cluster or spontaneously fissionable nucleus is moving (Fig.1). This is a positive potential of Coulomb repulsion $V_{Coul}$ out of nucleus ($r>R$) and a negative potential of nuclear attraction $V_{nucl}$, for example, of rectangular form, within nucleus ($r<R_{nucl}$). Note that the Kramers velocity (14) depends only on barrier height and curvature of potential in its extremes, therefore the exact shape of potential is inessential. It in full measure concerns the exact form of nuclear attraction potential. Therefore it is important to note that obtained below results can be qualitatively applied to a wide class of bistable systems.

It is well known that it is possible to determinate approximately the nuclear radii from the experimental data of α-decay. The proximity means that in such experiments the distance between the centers of nucleus and α-particle, where nuclear forces cease to act, are measured. In other words, the distance $R$ equal to the sum of nuclear radius $R_{A-4,Z-2}$, α–particle radius $R_\alpha$ and nuclear force action radius $R_{nf}$ is determined. Thus, the Coulomb repulsion potential out of nucleus is acting at distances $r>R=R_{A-4,Z-2}+ R_\alpha+ R_{nf}$. The same is true for the Coulomb interaction radius in case of cluster radioactivity and spontaneous fission. At the same time the probability effective current $w_K$ over the Coulomb potential barrier by virtue of geometry (Fig. 1) and Eq. (14) will be described by the following simplified expression:

$$w_K = \frac{b}{2\pi} \exp\left(-\frac{V_{Coul} - E_{TKE}}{\langle \varepsilon \rangle}\right), \tag{16}$$

where $E_{TKE}=E_\alpha \approx Q_X$, $E_{TKE}=E_{cl}$ or $E_{TKE}=E_{SF}$ is the decay kinetic energy for α–decay, cluster radioactivity (cl) or spontaneous fission (SF), respectively; $Q_X$ is total decay energy.

We suppose below that the excitation heat energy $E^*$ is some part of the decay kinetic energy $E_{TKE}$:

$$E^* = \mu E_{TKE}, \quad \mu << 1. \tag{17}$$

Physical sense and the explanation of the necessity of this condition we will give below.

Substituting Eq. (17) into Eq.(15) and after that the obtained result into Eq.(14), we have the probability effective current $w_K$ (16) over potential barrier:

$$w_K = \frac{\langle \omega \rangle_{Kramers}}{2\pi} \exp\left[-\left(\frac{A}{8\mu}\right)^{1/2} \frac{V_{Coul} - E_{TKE}}{\sqrt{E_{TKE}}}\right]. \tag{18}$$

Then by virtue of equality $T_K=T_{1/2}=(w_K)^{-1}$ we find the Kramers effective time

$$\lg T_{1/2} = -\lg \frac{\langle \omega \rangle_{Kramers}}{2\pi} + \lg e \left( \frac{A}{8\mu} \right)^{1/2} \frac{V_{Coul} - E_{TKE}}{\sqrt{E_{TKE}}}, \tag{19}$$

where $T_{1/2}$ is half-life; $\langle \omega \rangle_{Kramers}$ is the effective frequency of $\alpha$–particle appearance on the nuclear surface of radius $R$;

$$V_{Coul} = \frac{(Z - Z_{cl})z_{cl}}{R_{Coul}} = \frac{(Z - Z_{cl})z_{cl}}{R_{A-A_{cl}, Z-Z_{cl}} + R_{cl} + R_{nf}}, \quad [MeV], \tag{20}$$

where $A$ and $Z$ are mass number and the charge of parent nucleus; $Z_{cl}$ is the charge of outgoing particle; $(Z-Z_{cl})$ is the charge of the daughter nucleus; $R_{Coul}$ is minimal Coulomb radius [Fm].

Now it is easy to explain the necessity of assumption (17). It is stipulated by the fact that just under this condition the expression for Kramers effective time (19) can be represented by following approximate formula:

$$\lg T_{1/2} \cong \frac{C}{\sqrt{E_{TKE}}} - B, \tag{21}$$

which is one of the variants of the experimentally established at the earliest stage of nuclear physics Geiger-Nuttall law for $\alpha$–decay. In 1989-1990 such an experimental dependence was also discovered for cluster radioactivity [27, 28].

To explain the experimental law (21) predetermining the large variations of the half-life of heavy nuclei, in 1928 the theory of quantum-mechanical tunneling of α- particle over the Coulomb barrier was proposed. This mechanism in the framework of Gamow theory [1] is reduced to the following expression for the passage time (the half-life $T_{1/2}$ of heavy nuclei) of $\alpha$–particle through potential barrier:

$$\lg T_{1/2} = -\lg \frac{\langle \omega \rangle_{Gamov}}{2\pi} + \lg e \frac{4e^2(Z-2)}{\hbar} \sqrt{\frac{2\mu_\alpha}{E_\alpha}} \left[ Arc \cos \sqrt{x} - \sqrt{x(1-x)} \right], \quad x = \frac{R_{Col}}{r_T}, \tag{22}$$

where $\mu_\alpha = (A-4)4/A$ is reduced mass; $\hbar$ is reduced Planck constant, $\langle \omega \rangle_{Gamov}$ is effective frequency of $\alpha$–particle appearance on nuclear surface of radius $R_{Coul}$; $r_T = e^2(Z-2)2/E_\alpha$.

If the Coulomb barrier height is much greater than energy $E_\alpha$, what is typical situation for all natural $\alpha$–radiators (i.e. $x = R_{Coul}/r_T \ll 1$), and the term in squared brackets in Eq. (22) is approximately equal to $(0.5\pi - 2x^{-1/2})$, Eq. (22) for half-life is simplified and looks like:

$$\lg T_{1/2} = -\lg \frac{\langle \omega \rangle_{Gamov}}{2\pi} + \lg e \frac{\pi e^2 2(Z-2)}{\hbar} \sqrt{\frac{2\mu_\alpha}{E_\alpha}} - \lg e \frac{2}{\hbar} \sqrt{\mu_\alpha e^2 2(Z-2) R_{Col}}. \tag{23}$$

It is not hard to show that Eqs. (19) and (23) characterizing the "Kramers's" and "Gamow's" half-life, respectively, are equally well described by relation of Geiger-Nuttall type (21). However in both cases we do not have any information concerning the effective frequencies ($\langle \omega \rangle_{Gamov}$ and $\langle \omega \rangle_{Kramers}$) of particle-cluster appearance on the nuclear surface of radius $R_{Col}$ just as concerning the value of this radius. In the "Kramers's" case the uncertainty of value $\mu$ is else added (see Eq.(19)). Note that due to the uncertainty of effective frequency of α-particle appearance on nuclear surface of radius $R_{Col}$, we obtain by Eq. (23) only the order of half-life rather than its exact value. The second uncertainty can be explained by the fact that, as it is evident from the quadrupole moment measurements, the majority of α-radioactive nuclei are not spherical as it was supposed by Gamow α-decay theory, but they have ellipsoidal form with a ratio of longer to shorter half-axes running up to 1.5. Since the barrier penetrability of non-spherical nucleus varies in different places and is especially high near the "ends" of a nucleus, the estimations of nuclei radius obtained from alpha-decay data give the overestimated values which characterize not a certain effective radius, but actually longitudinal radius of a nucleus.

Thus, the dependence on decay energy and Coulomb barrier characteristics for cluster radioactivity and spontaneous fission has the same nature as alpha-decay, the problem of the indicated uncertainties remains. It specially concerns the uncertainty of effective frequencies ($\langle\omega\rangle_{Gamov}$ and $\langle\omega\rangle_{Kramers}$) of cluster-particle appearance on a nuclear surface of radius $R_{Coul}$, as the corresponding theoretical estimations are extremely difficult to obtain, and even if they are obtained, they are very approximate [29,30].

At the same time, the energy, half-life and relative decay probabilities for the majority of radioactive heavy nuclei are well measured within the framework of alpha-, cluster- and fission-fragment spectroscopy. These data are collected in the well-known ENSDF nuclear data library [31] and in a combination with theoretical estimations (19) make it possible to solve one of the primary problems of nuclear spectroscopy of decay processes, which is formulated in the following way. Using the experimental ENSDF data, for example, $T_{1/2}^{\exp}$, and theoretical estimations $T_{1/2}^{theory}$ (see Eqs. (17) and (21)) it is necessary to solve the inverse nonlinear problem, which can be described by the system of nonlinear equations

$$T_{1/2}^{\exp} = T_{1/2}^{theory}(E_{TKE}, A, Z, A_{cl}, Z_{cl}, R_{Coul}, \omega, \mu). \qquad (24)$$

with respect to unknown parameters $R_{Coul}$, $\omega$ and $\mu$.

The solution of the system of nonlinear algebraic equations of such a type under certain conditions allows to obtain a set of important data on intranuclear processes. In particular, it makes possible to obtain a functional dependence of effective frequency $\langle\omega\rangle_{Kramers}$) of cluster-particle occurrence on a nuclear surface of radius $R_{Coul}$, and also the dependence of radius $R_{Coul}$ on quantum numbers (in our case, mass number and charge) characterizing the parent and daughter nuclei. On the other hand, the large variations of half-lifes will lead to situation, when common determinant of the system will have many "zeros", and as a whole the system will be quasi-degenerate. This means that we have the ill-conditioned system of nonlinear equations, whose solutions can be instable to the low changes of initial data. In other words, a problem of this type belongs to a class of ill-posed problems, and to solve it we used the Alexandrov dynamic autoregularization method (FORTRAN code REGN-Dubna [32]) which is constructive development of Tikhonov regularization method [33].

We present below the results of solving the inverse nonlinear problem in the framework of Kramers (19) unified description of stochastic channels for α–decay, cluster radioactivity and spontaneous fission by the Alexandrov dynamic autoregularization method, using the well-known experimental data.

## 3. Comparing theory with experiment

In the case of heavy nucleus radioactive decay with heavy cluster emission (such as $^{14}C$, $^{24}Ne$, $^{28}Mg$, $^{34}Si$) as well as in the case of α-decay, the inequality $Q_{cl} < B_{A_{cl}(A-A_{cl})}^{Coul}$ is fulfilled, where $Q_{cl}$ is cluster decay energy and $B_{A_{cl}(A-A_{cl})}^{Coul}$ is Coulomb interaction energy between daughter nucleus (with mass number $A-A_{cl}$ and charge $Z-Z_{cl}$) and cluster (with mass number $A_{cl}$ and charge $Z_{cl}$) in contact point [28]. In other words, such a process is deep-subbarrier. Tacking into account, that experiment [28, 34] shows that the kinetic energy of decay products emitted from parent nucleus ($A$) remains almost the same:

$$E_{cl}^{\exp} \cong Q_{cl} \frac{A - A_{cl}}{A}, \qquad (25)$$

we can assume, that daughter nuclei and clusters are almost unexcited. Both of these arguments indicate there may be no noticeable parent nucleus reorganization during decay. Hence it is possible to assume that radioactive decay with heavy cluster emission is α-decay analogue [28]. In this sense the experimental result detecting the fine structure of cluster decay of $^{233}Ra$ [35] is very important. It is known, that this result was theoretically predicted in the framework of model [36] based on analogues with α-decay, and therefore this experiment, in fact gives decisive confirmation of analogy between mechanisms of α-decay and decay with heavy clusters emission [28].

Before analysis of computational experiment results, one important and unexpected fact has to be noted, i.e. the solution of inverse problem in the framework of Kramers (19) universal description of α-decay and cluster radioactivity was absolutely sufficient to describe the spontaneous fission without any additional adjusting parameters. This suggests that for α-decay as well as for cluster radioactivity and spontaneous fission the inequality of $E_{TKE} \neq Q_\alpha$ type is true. At least, it fulfils in known α-experiments for heavy and superheavy nuclei [37]. It can be partially explained by the fact that in α-decay, where the transition happens in one of the excited states of finite nucleus or vice versa - from one of the excited states of parent nucleus, the energy of α-particles is always less or more, respectively, than normal. Running a few steps forward, we can assume that making allowance for this strict inequality ($E_{TKE} \neq Q_\alpha$) will lead to sharp decrease of the parameterization parameter numbers of functions $R_{Kramers}$, $<\omega>_{Kramers}$, $\mu$ on quantum numbers $A$, $Z$, $A_{cl}$, $Z_{cl}$, in Eq. (24).

Now let us consider the solving of inverse problem in framework of Kramers (19) universal description of α-decay, cluster radioactivity and spontaneous fission, where alpha-particle is considered as a smallest cluster. Using the Alexandrov's dynamic regularization method [32] for solving the inverse problem (19) on the set of experimental data (Tables 1-3) from the ENSDF [31], we obtained phenomenological functional dependencies of previously unknown parameters in framework of Kramers (19) universal description of α-decay, cluster radioactivity and spontaneous fission.

In this case, the system of nonlinear equations (24) looks like

$$\lg T_{1/2}^{\exp} = \lg T_{1/2}^{Kramers}(E_{cl}, A, Z, Z_{cl}, R_{Kramers}(A, Z, A_{cl}, Z_{cl}), \omega_{Kramers}(R_{Kramers}), \mu(Z, A_{cl}, Z_{cl})), \qquad (26)$$

where we have applied the parameterization of functions $R_{Kramers}$, $\omega_{Kramers}$, $\mu$ with respect to quantum numbers $A$, $Z$, $A_{cl}$, $Z_{cl}$ (which determine the mass numbers and charges of parent nucleus and cluster) and energies $E_{TKE}$, $Q_{cl}$ (which determine the decay kinetic and total energy) in the following form:

$$\lg \frac{\langle \omega \rangle_{Kramers}}{2\pi} = a_{20} + \frac{1}{R_{Kramers}}, \qquad (27)$$

$$\mu = \exp\left[a_1 + a_2 \frac{(A-2Z)^2}{A^2} + a_3 \frac{A-A_{cl}}{A}\left(1 - \frac{E_{TKE}}{Q_{cl}}\right) + \left(a_4 \frac{A-A_{cl}}{A} + a_5 \frac{1}{Z_{cl}}\right)\left(1 - \frac{1}{Z_{cl}}\right)\right], \qquad (28)$$

$$R_{Kramers} = \left[B_1(A-Z_{cl})^{1/3} + B_1 A_{cl}^{1/3} - 1\right]B_2, \quad [fm], \qquad (29)$$

$$B_1 = \exp\left[a_6\left(\frac{A-2Z}{A}\right)^2 + a_7\frac{Z}{A} + \left(a_8 + a_9\frac{A-A_{cl}}{A} + a_{10}\frac{1}{A_{cl}}\right)\left(1-\frac{E_{TKE}}{Q_{cl}}\right) + \right.$$
$$\left. + \left(a_{11} + a_{12}\frac{A-A_{cl}}{A} + a_{13}\frac{1}{Z_{cl}}\right)\left(1-\frac{1}{Z_{cl}}\right)\right], \tag{30}$$

$$B_2 = \exp\left[a_{14}\frac{1}{Z} + a_{15}\left(\frac{A-2Z}{A}\right)^2 + a_{16}\frac{Z}{A} + a_{17}\frac{A-A_{cl}}{A}\left(1-\frac{E_{TKE}}{Q_{cl}}\right) + \right.$$
$$\left. + \left(a_{18} + a_{19}\frac{Z-Z_{cl}}{Z}\right)\left(1-\frac{1}{Z_{cl}}\right)\right]. \tag{31}$$

Note, one of ways for finding the hidden dependences of task parameters on characteristic variables (in our case the quantum numbers $A$, $Z$, $A_{cl}$, $Z_{cl}$ and energies $E_{TKE}$, $Q_{cl}$), which determine the state of investigated system, is briefly described in Ref. [38].

The solutions of inverse non-linear problem of Kramers type (26) for the ENSDF experimental data set (Table 1-3), which are presented as values of parameters $a_i$ and their relative errors $\Delta a_i / a_i$, are collected in the Table 4. Data in Tables 1-3 and Fig.2 show a good coincidence ($\chi^2$/NDF=82.5/72) of experimental and theoretical half-lifes of α-decay, cluster ($^{14}$C, $^{24}$Ne, $^{28}$Mg, $^{34}$Si) radioactivity and spontaneous fission depending on decay total kinetic energy $E_{TKE}$.

For the verification of obtained solution of inverse problem in the framework of Kramers universal description (19) of α–decay, cluster radioactivity and spontaneous fission, whose parameters are given in Table 4, we have used the experimental data of α–decay for superheavy nucleus, which were kindly given by Yu.Ts. Oganesian (JINR, Dubna, Russia) [37]. Fig.3 displays good accordance between experimental and theoretical (Kramers) half-lifes for alpha decay, depending on the decay energy $E_\alpha$. We consider that some lack of the coincidence of theoretical and experimental data in Fig.3 (see Table 4) is caused by low number of measurements (due to understandable reasons), which did not exceed 25 measurements for each decay type [37].

Basing on the solution of inverse problem in the framework of the Kramers (19) universal description of α–decay, cluster radioactivity and spontaneous fission, whose parameters are shown in Table 4, we give predictable value of energy $E_{TKE}$ for U$^{234}$ and Th$^{228}$ nuclei, which inclined to O$^{20}$ and Ne$^{24}$ cluster-radioactivity, respectively (see Table 2 and Fig.2). These data can be of interest for future experiments.

Finally, it is possible to conclude that received results are an indirect confirmation of the applicability of the Langevin fluctuation-dissipative dynamics and, in particular, of the Kramers diffusion mechanism [24-26] for the effective description of collective motions in nuclei generating the stochastic channel of α–decay, cluster radioactivity and spontaneous fission. Although a situation is, at first sight, complicated by the fact that dissipation in nucleus (or more precisely, nuclear friction) is magnitude unobserved experimentally [39], but, it turned out, there is the obvious possibility of the unique proof of its real existence. For example, it is well known, that the introduction of the external periodic signal into Langevin equation (12)-(13) must result in the observation of stochastic resonance [40]. In other words, the experiments on the search of nuclear stochastic resonance in α-decay, which can not in principle take place within the framework of probabilistic interpretation of quantum mechanics but must be observed within the framework of the Bohmian mechanics supplemented with the Chetaev's generalized theorem [12] can become the determinative factor for the revelation of fundamental role of dissipation not only in nuclear dynamics but in quantum physics generally.

At last, note that, when first researcher UV-lasers of frequency about $10^{18}$–$10^{20}$ $s^{-1}$ [41] will appear in the near future, the problem of the stochastic mode excitation in the atomic nucleus under action of periodic external field will become actual not only in respect to the direct study of dissipation and consequently of self-organization and quantum chaos in it, but fundamental in the view of possible break-through of "probabilistic smokescreen" [11] to the holistic understanding of causal interpretation of quantum physics satisfying the Einstein's locality principle.

**Acknowledgment**

The authors are grateful to L. Alexandrov (Bogolubov Laboratory of Theoretical Physics, JINR, Dubna) for granting REGN- Dubna software package.

**Table 1.** The ENSDF experimental data for α-decay of even-even nuclei and the theoretical half-lifes $T_{1/2}^{Theory}$ obtained by our model.

| No | Nuclei | A | Z | $A_{cl}$ | $Z_{cl}$ | $E_{TKE}$, Mev | $Q_\alpha$, Mev | $T_{1/2}^{Theory}$, year | $T_{1/2}^{Exprt}$, year | $\Delta T_{1/2}^{Exprt}$, year |
|---|---|---|---|---|---|---|---|---|---|---|
| 1 | Pt | 168 | 78 | 4 | 2 | 6.832±0.010 | 6.999±0.001 | 7.1228E-11 | 6.3376E-11 | 3.17E-12 |
| 2 | Pt | 174 | 78 | 4 | 2 | 6.038±0.004 | 6.184±0.001 | 3.6504E-08 | 2.8171E-08 | 5.39E-11 |
| 3 | Pt | 176 | 78 | 4 | 2 | 5.753±0.003 | 5.887±0.000 | 2.5613E-07 | 1.9963E-07 | 1.58E-08 |
| 4 | Pt | 178 | 78 | 4 | 2 | 5.446±0.003 | 5.561±0.000 | 9.1244E-07 | 6.6862E-07 | 1.90E-08 |
| 5 | Hg | 174 | 80 | 4 | 2 | 7.067±0.006 | 7.233±0.001 | 6.9551E-11 | 6.0207E-11 | 1.27E-12 |
| 6 | Hg | 180 | 80 | 4 | 2 | 6.119±0.004 | 6.258±0.000 | 6.9882E-08 | 8.1755E-08 | 3.17E-10 |
| 7 | Hg | 182 | 80 | 4 | 2 | 5.867±0.005 | 5.999±0.001 | 6.1335E-07 | 3.4318E-07 | 1.90E-09 |
| 8 | Pb | 186 | 82 | 4 | 2 | 6.332±0.007 | 6.471±0.001 | 7.4096E-08 | 1.5305E-07 | 1.58E-09 |
| 9 | Pb | 188 | 82 | 4 | 2 | 5.983±0.004 | 6.111±0.000 | 1.0762E-06 | 7.9537E-07 | 3.17E-10 |
| 10 | Po | 188 | 84 | 4 | 2 | 7.910±0.013 | 8.082±0.001 | 7.0991E-12 | 9.5064E-12 | 9.51E-13 |
| 11 | Po | 190 | 84 | 4 | 2 | 7.537±0.006 | 7.699±0.001 | 6.8357E-11 | 7.7636E-11 | 1.58E-12 |
| 12 | Po | 192 | 84 | 4 | 2 | 7.167±0.007 | 7.322±0.001 | 1.0062E-09 | 1.0520E-09 | 4.44E-12 |
| 13 | Po | 194 | 84 | 4 | 2 | 6.843±0.003 | 6.990±0.000 | 1.0917E-08 | 1.2422E-08 | 1.27E-14 |
| 14 | Po | 196 | 84 | 4 | 2 | 6.520±0.023 | 6.657±0.000 | 1.1203E-07 | 1.8157E-07 | 7.29E-10 |
| 15 | Po | 198 | 84 | 4 | 2 | 6.182±0.022 | 6.309±0.002 | 1.5956E-06 | 3.3653E-06 | 5.70E-09 |
| 16 | Po | 200 | 84 | 4 | 2 | 5.862±0.018 | 5.981±0.002 | 2.9636E-05 | 2.1865E-05 | 1.90E-08 |
| 17 | Po | 202 | 84 | 4 | 2 | 5.588±0.017 | 5.686±0.002 | 4.9223E-05 | 8.4987E-05 | 9.51E-08 |
| 18 | Po | 206 | 84 | 4 | 2 | 5.224±0.015 | 5.327±0.001 | 2.9137E-02 | 2.4093E-02 | 2.74E-05 |
| 19 | Po | 214 | 84 | 4 | 2 | 7.687±0.007 | 7.849±0.001 | 4.1992E-12 | 5.2064E-12 | 6.34E-15 |
| 20 | Po | 216 | 84 | 4 | 2 | 6.778±0.005 | 6.906±0.001 | 2.0487E-09 | 4.5948E-09 | 6.34E-12 |
| 21 | Po | 218 | 84 | 4 | 2 | 6.002±0.009 | 6.115±0.001 | 3.9984E-06 | 5.8940E-06 | 3.80E-09 |
| 22 | Rn | 198 | 86 | 4 | 2 | 7.205±0.005 | 7.349±0.000 | 1.9900E-09 | 2.0597E-09 | 9.51E-12 |
| 23 | Rn | 200 | 86 | 4 | 2 | 6.902±0.003 | 7.043±0.000 | 2.8424E-08 | 3.0421E-08 | 9.51E-10 |
| 24 | Rn | 202 | 86 | 4 | 2 | 6.640±0.019 | 6.774±0.002 | 2.1262E-07 | 3.1688E-07 | 9.51E-10 |
| 25 | Rn | 208 | 86 | 4 | 2 | 6.140±0.017 | 6.271±0.002 | 7.0718E-05 | 4.6296E-05 | 2.66E-08 |
| 26 | Rn | 218 | 86 | 4 | 2 | 7.129±0.012 | 7.263±0.002 | 1.1089E-09 | 1.1091E-09 | 1.58E-11 |
| 27 | Rn | 220 | 86 | 4 | 2 | 6.288±0.010 | 6.405±0.001 | 2.2059E-06 | 1.7619E-06 | 3.17E-10 |
| 28 | Rn | 222 | 86 | 4 | 2 | 5.489±0.030 | 5.590±0.000 | 1.3687E-02 | 1.0468E-02 | 8.21E-08 |
| 29 | Ra | 204 | 88 | 4 | 2 | 7.486±0.006 | 7.636±0.001 | 2.1582E-09 | 1.8062E-09 | 3.49E-12 |
| 30 | Ra | 212 | 88 | 4 | 2 | 6.899±0.017 | 7.040±0.002 | 3.3566E-07 | 4.1195E-07 | 6.34E-09 |
| 31 | Ra | 214 | 88 | 4 | 2 | 7.137±0.003 | 7.283±0.000 | 4.7379E-08 | 7.7953E-08 | 9.51E-11 |
| 32 | Ra | 220 | 88 | 4 | 2 | 7.453±0.007 | 7.592±0.001 | 7.6376E-10 | 5.7039E-10 | 6.34E-13 |
| 33 | Ra | 222 | 88 | 4 | 2 | 6.559±0.005 | 6.679±0.000 | 1.3684E-06 | 1.1462E-06 | 3.17E-10 |
| 34 | Ra | 224 | 88 | 4 | 2 | 5.685±0.015 | 5.789±0.000 | 1.3964E-02 | 1.0021E-02 | 1.10E-05 |
| 35 | Ra | 226 | 88 | 4 | 2 | 4.784±0.025 | 4.871±0.000 | 1.9239E+03 | 1.6000E+03 | 7.00E-01 |
| 36 | Th | 210 | 90 | 4 | 2 | 7.899±0.017 | 8.053±0.002 | 6.1465E-10 | 2.8519E-10 | 5.39E-11 |
| 37 | Th | 216 | 90 | 4 | 2 | 7.922±0.008 | 8.081±0.001 | 7.1259E-10 | 8.2389E-10 | 4.75E-11 |
| 38 | Th | 218 | 90 | 4 | 2 | 9.666±0.010 | 9.849±0.001 | 4.2241E-15 | 3.4540E-15 | 4.12E-17 |
| 39 | Th | 222 | 90 | 4 | 2 | 7.980±0.002 | 8.127±0.001 | 1.1147E-10 | 7.0981E-11 | 4.12E-14 |
| 40 | Th | 224 | 90 | 4 | 2 | 7.170±0.010 | 7.304±0.001 | 7.4860E-08 | 3.3272E-08 | 6.34E-11 |
| 41 | Th | 226 | 90 | 4 | 2 | 6.337±0.010 | 6.444±0.001 | 3.7302E-05 | 5.8122E-05 | 1.90E-08 |
| 42 | Th | 228 | 90 | 4 | 2 | 5.423±0.022 | 5.520±0.002 | 2.6517E+00 | 1.9120E+00 | 2.00E-03 |
| 43 | Th | 230 | 90 | 4 | 2 | 4.687±0.015 | 4.770±0.002 | 6.9160E+04 | 7.5386E+04 | 3.00E+02 |
| 44 | Th | 232 | 90 | 4 | 2 | 4.012±0.014 | 4.083±0.001 | 9.5977E+09 | 1.4050E+10 | 6.00E+07 |
| 45 | U | 226 | 92 | 4 | 2 | 7.570±0.020 | 7.704±0.001 | 1.2414E-08 | 1.1091E-08 | 4.75E-09 |
| 46 | U | 228 | 92 | 4 | 2 | 6.680±0.010 | 6.796±0.001 | 1.9017E-05 | 1.7302E-05 | 3.80E-08 |
| 47 | U | 230 | 92 | 4 | 2 | 5.888±0.007 | 5.993±0.001 | 1.0152E-01 | 5.6947E-02 | 5.75E-04 |
| 48 | U | 232 | 92 | 4 | 2 | 5.320±0.014 | 5.414±0.001 | 9.5117E+01 | 6.8890E+01 | 4.00E-02 |
| 49 | U | 234 | 92 | 4 | 2 | 4.775±0.014 | 4.858±0.001 | 1.7898E+05 | 2.4549E+05 | 6.00E+01 |
| 50 | U | 236 | 92 | 4 | 2 | 4.494±0.003 | 4.573±0.001 | 2.8904E+07 | 2.3421E+07 | 4.00E+03 |
| 51 | U | 238 | 92 | 4 | 2 | 4.198±0.003 | 4.270±0.001 | 4.4627E+09 | 4.4680E+09 | 3.00E+05 |
| 52 | Pu | 236 | 94 | 4 | 2 | 5.768±0.008 | 5.867±0.001 | 2.6875E+00 | 2.8580E+00 | 8.00E-04 |
| 53 | Pu | 238 | 94 | 4 | 2 | 5.499±0.020 | 5.593±0.002 | 8.2055E+01 | 8.7713E+01 | 1.00E-02 |
| 54 | Pu | 240 | 94 | 4 | 2 | 5.168±0.015 | 5.256±0.001 | 7.9892E+03 | 6.5610E+03 | 7.00E-01 |
| 55 | Pu | 242 | 94 | 4 | 2 | 4.902±0.009 | 4.984±0.001 | 3.6001E+05 | 3.7360E+05 | 1.10E+02 |
| 56 | Pu | 244 | 94 | 4 | 2 | 4.589±0.001 | 4.666±0.001 | 8.0242E+07 | 8.0012E+07 | 9.00E+04 |
| 57 | Cm | 238 | 96 | 4 | 2 | 6.520±0.050 | 6.608±0.004 | 2.4091E-04 | 2.7379E-04 | 1.14E-06 |
| 58 | Cm | 240 | 96 | 4 | 2 | 6.291±0.005 | 6.398±0.001 | 7.1683E-02 | 7.3922E-02 | 2.74E-04 |
| 59 | Cm | 242 | 96 | 4 | 2 | 6.113±0.008 | 6.216±0.000 | 4.9037E-01 | 4.4617E-01 | 1.64E-05 |
| 60 | Cm | 244 | 96 | 4 | 2 | 5.805±0.005 | 5.902±0.000 | 1.8000E+01 | 1.8100E+01 | 1.00E-02 |
| 61 | Cm | 246 | 96 | 4 | 2 | 5.387±0.010 | 5.475±0.001 | 3.0607E+03 | 4.7596E+03 | 4.00E+00 |
| 62 | Cm | 248 | 96 | 4 | 2 | 5.078±0.025 | 5.162±0.003 | 4.0888E+05 | 3.4800E+05 | 6.00E+02 |
| 63 | Cf | 240 | 98 | 4 | 2 | 7.590±0.010 | 7.719±0.001 | 2.9232E-06 | 1.8252E-06 | 2.85E-07 |
| 64 | **Cf** | 244 | 98 | 4 | 2 | 7.209±0.004 | 7.329±0.002 | 5.9464E-05 | 3.6885E-05 | 1.14E-07 |
| 65 | **Cf** | 246 | 98 | 4 | 2 | 6.750±0.010 | 6.862±0.001 | 4.6022E-03 | 4.0726E-03 | 5.70E-06 |
| 66 | **Cf** | 248 | 98 | 4 | 2 | 6.258±0.005 | 6.361±0.001 | 7.9071E-01 | 9.1293E-01 | 7.67E-04 |
| 67 | **Cf** | 250 | 98 | 4 | 2 | 6.030±0.020 | 6.128±0.002 | 1.0019E+01 | 1.3081E+01 | 9.00E-03 |
| 68 | **Cf** | 252 | 98 | 4 | 2 | 6.118±0.004 | 6.217±0.000 | 3.7474E+00 | 2.6450E+00 | 8.00E-04 |
| 69 | **Fm** | 246 | 100 | 4 | 2 | 8.237±0.015 | 8.361±0.002 | 2.8942E-08 | 3.4857E-08 | 6.34E-10 |
| 70 | **Fm** | 248 | 100 | 4 | 2 | 7.870±0.020 | 8.000±0.001 | 1.4652E-06 | 1.1408E-06 | 9.51E-09 |
| 71 | **Fm** | 250 | 100 | 4 | 2 | 7.436±0.012 | 7.558±0.001 | 5.3393E-05 | 6.2742E-05 | 5.70E-07 |
| 72 | **Fm** | 252 | 100 | 4 | 2 | 7.039±0.002 | 7.155±0.002 | 2.3293E-03 | 2.8964E-03 | 5.70E-06 |
| 73 | **Fm** | 254 | 100 | 4 | 2 | 7.192±0.002 | 7.308±0.002 | 3.8158E-04 | 3.6961E-04 | 2.28E-07 |
| 74 | **No** | 252 | 102 | 4 | 2 | 8.415±0.001 | 8.549±0.001 | 9.3074E-08 | 7.1932E-08 | 4.44E-10 |
| 75 | **No** | 256 | 102 | 4 | 2 | 8.448±0.001 | 8.581±0.001 | 4.8203E-08 | 9.2212E-08 | 1.58E-10 |
| 76 | **Sg** | 260 | 106 | 4 | 2 | 9.770±0.003 | 9.912±0.003 | 1.2100E-10 | 1.1408E-10 | 2.85E-12 |
| 77[*] | – | 294 | 118 | 4 | 2 | 11.650±0.060 | 11.838±0.060 | 2.8198E-11 | 2.8202E-11 | +3.39E-11 −8.17E-12 |
| 78[*] | – | 292 | 116 | 4 | 2 | 10.660±0.070 | 10.809±0.070 | 4.9622E-10 | 5.7039E-10 | +5.07E-10 −2.14E-10 |
| 79[*] | – | 290 | 116 | 4 | 2 | 10.840±0.080 | 10.990±0.080 | 1.8945E-10 | 2.2499E-10 | +1.01E-10 −1.20E-10 |
| 80[*] | – | 288 | 114 | 4 | 2 | 9.940±0.060 | 10.091±0.060 | 2.6095E-08 | 2.5350E-08 | +8.56E-9 −1.50E-8 |
| 81[*] | – | 286 | 114 | 4 | 2 | 10.190±0.060 | 10.339±0.060 | 3.8586E-09 | 4.1195E-09 | +1.27E-09 −2.06E-9 |

[*] Data is given by Yu.Ts. Oganesian [37].

**Table 2.** The ENSDF experimental data for cluster radioactivity of even-even nuclei and the theoretical half-lifes $T_{1/2}^{Theory}$ obtained by our model.

| No | Nuclei | A | Z | $A_{cl}$ | $Z_{cl}$ | $E_{TKE}$, Mev | $Q_X$, Mev | $T_{1/2}^{Theory}$, year | $T_{1/2}^{Exprt}$, year | $\Delta T_{1/2}^{Exprt}$, year |
|----|--------|-----|----|----|----|-----------------|---------------|-----------|-----------|-----------|
| 82 | Ra | 226 | 88 | 14 | 6  | 26.46±1.00  | 28.79±1.00  | 6.315E+13 | 6.32E+13 | 3.70E+13 |
| 83 | Ra | 224 | 88 | 14 | 6  | 28.63±1.00  | 30.53±1.00  | 1.833E+08 | 2.52E+08 | 8.01E+07 |
| 84 | Ra | 222 | 88 | 14 | 6  | 30.97±1.00  | 33.05±1.00  | 8.369E+03 | 3.17E+03 | 4.69E+02 |
| 85 | Th | 230 | 90 | 24 | 10 | 51.98±1.00  | 57.68±1.00  | 1.208E+17 | 1.26E+17 | 2.21E+16 |
| 86 | U  | 232 | 92 | 24 | 10 | 55.86±1.00  | 62.31±1.00  | 1.370E+13 | 1.00E+13 | 7.17E+11 |
| 87 | U  | 234 | 92 | 28 | 12 | 65.26±1.00  | 74.13±1.00  | 3.097E+18 | 1.59E+18 | 7.48E+16 |
| 88 | Pu | 236 | 94 | 28 | 12 | 70.22±1.00  | 79.60±1.00  | 1.139E+14 | 1.59E+14 | 1.58E+14 |
| 89 | Cm | 242 | 96 | 34 | 14 | 82.88±1.00  | 96.43±1.00  | 9.977E+13 | 1.00E+14 | 5.86E+13 |
| 90 | Th | 228 | 90 | 20 | 8  | 40.44±1.00* | 44.73±1.00  | 1.792E+13 | 1.68E+13 | - |
| 91 | U  | 234 | 92 | 24 | 10 | 51.80±1.00* | 58.84±1.00  | 1.869E+18 | 2.52E+18 | - |

* Predicted values.

**Table 3.** The ENSDF experimental data for spontaneous fission of even-even nuclei and the theoretical half-lifes $T_{1/2}^{Theory}$ obtained by our model.

| No | Nuclei | A | Z | $A_{cl}$ | $Z_{cl}$ | $E_{TKE}$, Mev | $Q_X$, Mev | $T_{1/2}^{Theory}$, year | $T_{1/2}^{Exprt}$, year | $\Delta T_{1/2}^{Exprt}$, year |
|-----|--------|-----|-----|-----|----|-------------|--------------|------------|----------|----------|
| 92  | U  | 236 | 92  | 94  | 37 | 165.0±1.0 | 181.49±1.00 | 2.5014E+16 | 2.00E+16 | 1.00E+15 |
| 93  | U  | 236 | 92  | 93  | 37 | 164.0±1.0 | 185.31±1.00 | 1.6000E+16 | 1.60E+16 | 2.00E+15 |
| 94  | Pu | 240 | 94  | 96  | 38 | 172.0±1.0 | 194.88±1.00 | 1.3400E+11 | 1.34E+11 | 2.00E+09 |
| 95  | Cm | 244 | 96  | 97  | 38 | 185.5±1.0 | 200.76±1.00 | 8.5405E+06 | 1.34E+07 | 2.00E+05 |
| 96  | Cm | 250 | 96  | 100 | 38 | 182.3±1.0 | 198.18±1.00 | 1.9154E+04 | 2.00E+04 | 5.00E+02 |
| 97  | Cf | 254 | 98  | 102 | 39 | 186.1±1.0 | 206.72±1.00 | 1.8165E-01 | 1.78E-01 | 5.48E-04 |
| 98  | Cf | 252 | 98  | 101 | 39 | 186.5±1.0 | 207.73±1.00 | 8.5490E+01 | 8.55E+01 | 1.00E+00 |
| 99  | Cf | 246 | 98  | 98  | 39 | 195.6±1.0 | 209.03±1.00 | 2.3353E+03 | 2.00E+03 | 2.00E+02 |
| 100 | Fm | 258 | 100 | 103 | 40 | 200.3±1.0 | 220.90±1.00 | 3.8000E-11 | 3.80E-11 | 6.34E-13 |

**Table 4.** The values of parameters $a_i$ and their relative errors $\Delta a_i/a_i$.

| $i$ | $a_i$ | $\Delta a_i/a_i$, % |
|---|---|---|
| 1 | -0.5786501537235E+01 | 2.90 |
| 2 | -0.2096263480335E+02 | 1.90 |
| 3 | -0.3814591516659E+02 | 2.70 |
| 4 | 0.6900587198207E+01 | 2.70 |
| 5 | 0.7660345598675E+01 | 5.00 |
| 6 | 0.1908313257301E+02 | 1.40 |
| 7 | 0.1826397833295E+02 | 1.30 |
| 8 | 0.5919551276390E+01 | 1.60 |
| 9 | -0.1028171722816E+02 | 1.50 |
| 10 | -0.4411225968202E+02 | 3.50 |
| 11 | -0.1131089043128E+02 | 1.00 |
| 12 | 0.1314247777365E+01 | 1.20 |
| 13 | 0.2865440882262E+01 | 2.60 |
| 14 | 0.4240393211738E+01 | 18.00 |
| 15 | -0.1229614115313E+02 | 2.20 |
| 16 | -0.1772081454140E+02 | 1.20 |
| 17 | 0.1689691120764E+02 | 0.75 |
| 18 | 0.1666949024191E+02 | 0.82 |
| 19 | -0.8643631861055E+01 | 0.78 |
| 20 | 0.2749864919484E+02 | 1.20 |

**Figure captions**

**Fig. 1.** The dependence of nuclear particle potential energy on distance to the nuclear center.

**Fig. 2.** The theoretical and experimental values of half-life for even-even nuclei as function of the fission total kinetic energy $E_{TKE}$ for $\alpha$–decay, the cluster radioactivity and the spontaneous fission.

**Fig. 3.** The theoretical and experimental values of the half-life of even-even-nuclei as function of the fission total kinetic energy $E_{TKE}$ for $\alpha$–decay of superheavy nuclei with Z=114, 116, 118.

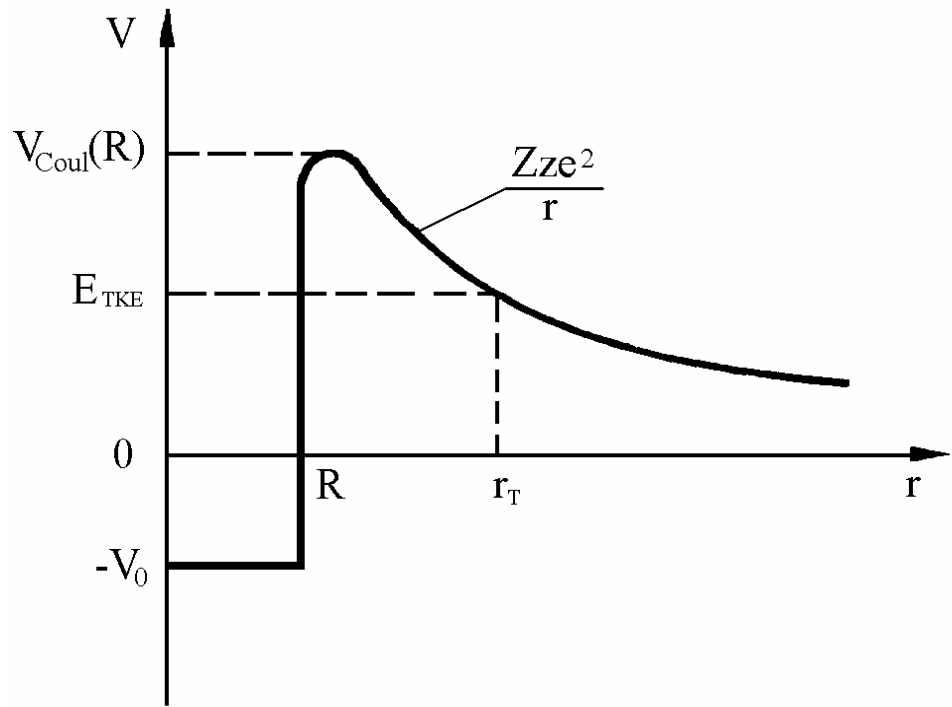

Fig. 1

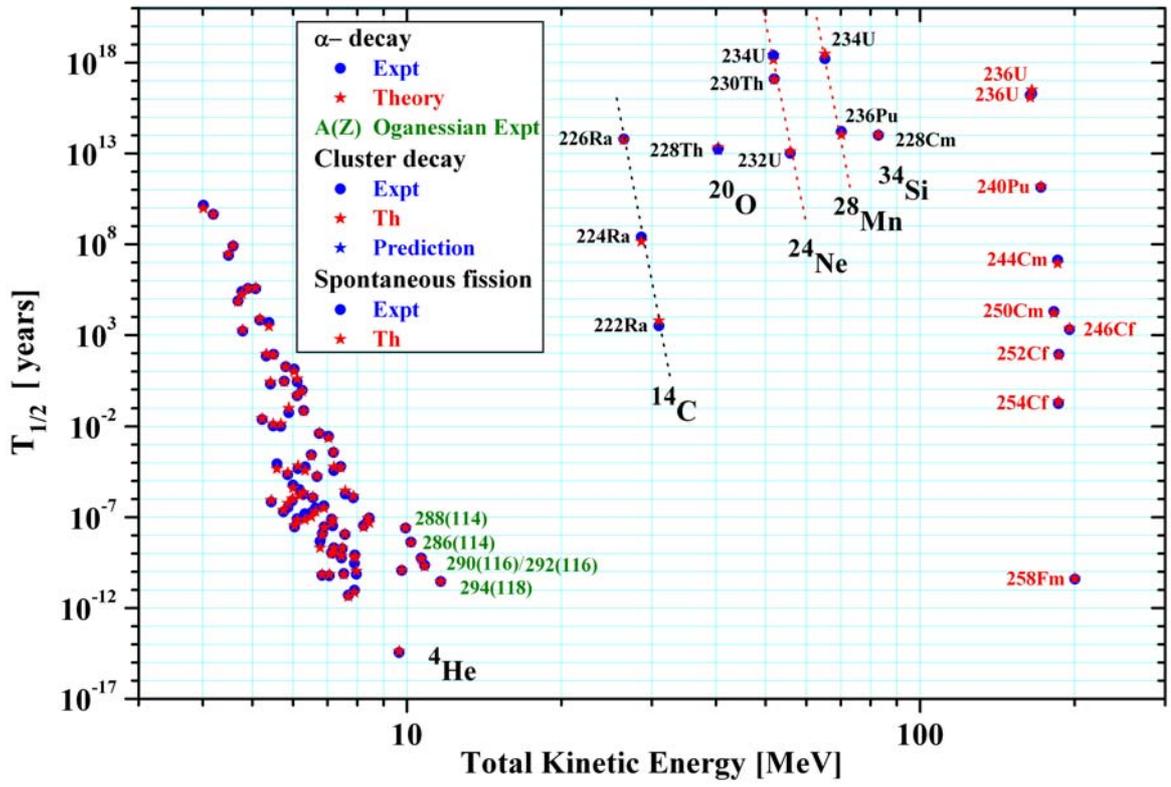

Fig. 2.

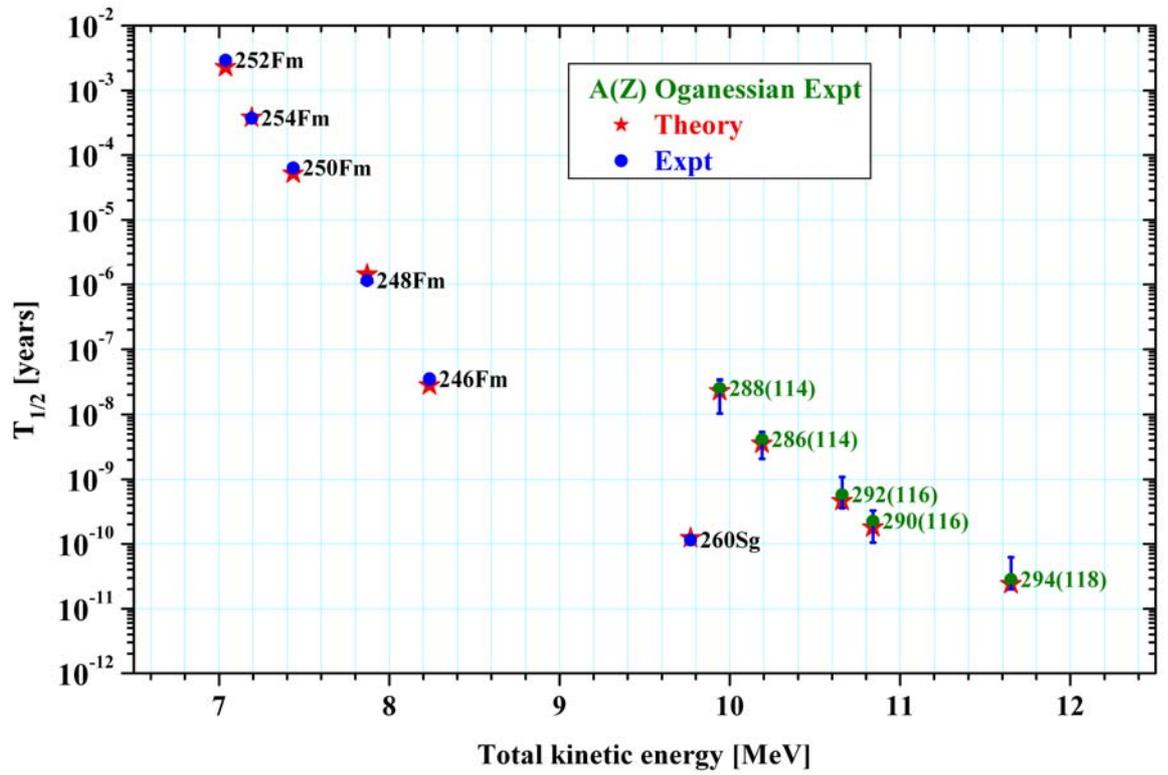

Fig. 3.